\documentstyle[aps,prb,eqsecnum,epsf,floats]{revtex}
\begin{document}
\def\SNG{{\em Physical Review Style and Notation Guide}}
\def\LUG {{\em \LaTeX{} User's Guide \& Reference Manual}}
\def\btt#1{{\tt$\backslash$\string#1}}%
\def\REVTeX{REV\TeX}
\def\AmS{{\protect\the\textfont2
        A\kern-.1667em\lower.5ex\hbox{M}\kern-.125emS}}
\def\AmSLaTeX{\AmS-\LaTeX}
\def\BibTeX{\rm B{\sc ib}\TeX}
\twocolumn[\hsize\textwidth\columnwidth\hsize\csname@twocolumnfalse%
\endcsname
\title{Quantum critical behavior of disordered itinerant ferromagnets}
\author{T.R.Kirkpatrick}
\address{Institute for Physical Science and Technology, and Department of Physics\\
University of Maryland,\\ 
College Park, MD 20742}
\author{D.Belitz}
\address{Department of Physics and Materials Science Institute\\
University of Oregon,\\
Eugene, OR 97403}

\date{\today}
\maketitle

\begin{abstract}
The quantum ferromagnetic transition at zero temperature in disordered
itinerant electron systems is considered. Nonmagnetic quenched
disorder leads to diffusive electron dynamics that induces an effective
long-range interaction between the spin or order parameter fluctuations
of the form $r^{2-2d}$, with $d$ the spatial dimension. This leads to
unusual scaling behavior at the quantum critical point, which is determined
exactly. In three-dimensional systems the quantum critical exponents are
substantially different from their finite temperature counterparts, a 
difference that should
be easily observable. Experiments to check these predictions are proposed.
\end{abstract}
]
\section{Introduction}
\label{sec:I}

The general problem of describing quantum phase transitions is a subject
of great current interest. These transitions occur at zero temperature as
a function of some nonthermal control parameter, and the relevant 
fluctuations are of quantum nature rather than of thermal origin. Early
work in this field established that if the quantum phase transition has a
classical analogue at finite temperature, then
in the physical dimension, $d=3$, the former tends to have a
simpler critical behavior than the latter.
In particular, one of the most obvious examples of a
quantum phase transition, namely the ferromagnet-to-paramagnet transition 
of itinerant electrons at zero temperature, $T=0$, as a function 
of the exchange interaction between the electron spins, was found to have
mean-field critical behavior in $d=3$.\cite{Hertz} 
The reason for this tendency is that
the coupling between statics and dynamics that is inherent to quantum
statistics problems effectively increases the dimensionality of the system
from $d$ to $d+z$, with $z$ the dynamical critical exponent. In the case
of {\it clean} itinerant quantum ferromagnets, $z=3$ in mean-field theory.
This appears to reduce the upper critical dimension $d_c^+$, which is the 
dimension above which mean-field theory yields the exact critical behavior, 
from $d_c^+=4$ in the classical case to $d_c^+=1$ in the quantum 
case.\cite{Hertz} However, this conclusion has recently been 
disputed,\cite{Sachdev,clean} and the critical behavior of clean 
itinerant ferromagnets in low-dimensional systems is currently under renewed 
investigation.\cite{cleantbp}

It has been known for some time that if one adds quenched, nonmagnetic disorder
to the system, then the
critical behavior at the quantum phase transition must be substantially
modified,\cite{Millis} contrary to earlier suggestions.\cite{Hertz} 
It is known that the correlation length exponent, 
$\nu$, must satisfy the inequality $\nu\geq 2/d$ in systems with quenched
disorder.\cite{Harris} The physical origin of this constraint is the
requirement that disorder induced fluctuations in the location of the 
critical point in parameter space must be small compared to the distance 
from the critical point in order for the phase transition to be sharp.
Any mean-field, and any standard Gaussian theory
yields $\nu=1/2$, which is incompatible with this lower bound in any
dimension $d<4$. Technically, this implies that the disorder is a relevant
perturbation with respect to the mean-field or Gaussian fixed point.

From a perspective that is entirely focused on the statistical mechanics
aspects of the phase transition problem (and that therefore does not
take into account from the beginning the aspects that have to do with the 
underlying disordered electron system), it would be tempting to model
this disordered quantum phase transition by making the mass term in the
effective field theory for the clean case (i.e. the coefficient of the
term quadratic in the order parameter field in the Landau-Ginzburg-Wilson
(LGW) functional) a random variable in order to describe the fluctuations in
the location of the critical point. For the quantum 
paramagnet-to-ferromagnet transition under consideration one can
readily convince oneself that for $d<4$ the resulting disorder term is 
relevant in the renormalization group sense
with respect to the clean Gaussian fixed point. Presumably, the presence of
this relevant operator leads to a new critical fixed point with a
correlation length exponent $\nu\geq 2/d$.

It turns out, however, that such a model is not a technically appropriate
description of the quantum ferromagnetic transition in a system of
disordered itinerant electrons. In order to explain this important point,
let us anticipate a number of results that will be discussed in detail
in Sec.\ \ref{sec:II} below. If one attempts to derive the LGW functional
mentioned above, then the coefficient of the random mass term is a
correlation function whose expansion in the random potential is very
singular. The reason is that the dynamics of a quantum particle in a
disordered environment in the limit of small wavenumbers, $k$, and low
frequencies, $\omega$, are qualitatively different from those in a
clean system. In the former case the motion is diffusive, while in the
latter it is ballistic. Technically, the limit $k, \omega\rightarrow 0$
does not commute with the clean limit. To overcome this problem it is
advantageous to not expand in powers of the random potential, but to
expand instead in the fluctuations of the coefficients of the LGW
functional at fixed disorder.

This procedure, by construction, automatically resums the most obvious
singularities in the disordered itinerant electron problem. Although it will
turn out that the coefficients in the resulting LGW functional
are still singular, these remaining singularities can be handled
mathematically as we will see below. These singularities arise due to
what are often called weak localization effects.\cite{LeeRama,R}
Their physical origin lies in the fact that the dynamics of diffusive
electrons are intrinsically long-ranged. Via mode-coupling effects, this
feature couples back to the ferromagnetic transition problem. As we show in
this paper, the net
effect is that the interactions between the spin fluctuations in the
LGW theory are of long range. The critical behavior described by
the resulting nonlocal field theory can be determined exactly for all
$d>2$, and satisfies $\nu\geq 2/d$ as required. The critical exponents
obtained from this theory are $d$-dependent for all $d<6$. In $d=3$,
they are substantially different from either the mean-field exponents,
or from those for a classical Heisenberg model, which has striking
observable consequences.

The outline of this paper is as follows. In Sec.\ \ref{sec:II} we first
define an itinerant disordered electron model, and then discuss how to
derive an order parameter description for a ferromagnetic phase transition
starting from a fermionic field theory. We also discuss in detail the
coefficients in the LGW functional and show that they have long-range
properties. In Sec.\ \ref{sec:III} the critical behavior is determined
exactly for dimensions $d\geq 2$. For $d>6$, mean-field exponents are
obtained, while for $2<d<6$, $d$-dependent exponents are found. In Sec.\ 
\ref{sec:IV} the results of this paper are reviewed. First we summarize
the theoretical aspects of our results, and then 
we point out some of the experimental consequences we
expect.
In Appendix\ \ref{app:A} we calculate the wavenumber
dependent spin susceptibibility for disordered interacting fermions, and
in Appendix\ \ref{app:B} we discuss the logarithmic corrections
to scaling that exist at the critical dimensions $d_1^+=6$ and $d_2^+=4$,
as well as for all $2<d<4$.

\section{Model, and Theoretical Framework}
\label{sec:II}

In the first part of this section we define a simple model for interacting
electrons in a disordered environment. Starting with this general fermionic
field theory we then derive an LGW or order parameter description, 
with the spin
density fluctuation as the order parameter. In the last part of this
section we derive and discuss the coefficients in this LGW functional.
As mentioned in the introduction, the crucial point is that the interactions
in the effective LGW theory are long-ranged due to the diffusive dynamics
of the electrons in a disordered metal.

\subsection{The Model}
\label{subsec:II.A}

The partition function of any fermionic system can be written in the 
form,\cite{NegeleOrland}
\begin{mathletters}
\label{eqs:2.1}
\begin{equation}
Z=\int D{\bar\psi}\,D\psi\ \exp\left(S\left[{\bar\psi},\psi\right]\right)\quad,
\label{eq:2.1a}
\end{equation}
where the functional integration measure is defined with respect to 
Grassmannian (i.e., anticommuting) fields $\bar\psi$ and $\psi$, and $S$
is the action,
\begin{equation}
S = \int_0^{\beta} d\tau \int d{\bf x}\ 
   {\bar\psi}^i({\bf x},\tau)\,{\partial\over\partial\tau}\,\psi^i({\bf x},\tau)
 - \int_0^{\beta} d\tau\ H(\tau)\quad.
\label{eq:2.1b}
\end{equation}
\end{mathletters}%
Here ${\bf x}$ denotes positions and $\tau$ imaginary time, $H(\tau)$ 
is the Hamiltonian in imaginary time representation,
$\beta=1/T$ is the inverse temperature, $i=1,2$ denotes spin labels, and
summation over repeated indices is implied. Throughout this paper we use
units such that $k_B = \hbar = e^2 = 1$. Our starting model is a fluid
of interacting electrons moving in a static random potential $v({\bf x})$,
\begin{mathletters}
\label{eqs:2.2}
\begin{eqnarray}
H(\tau) = \int d{\bf x}\ \left[{1\over 2m}\,\nabla {\bar\psi}^i({\bf x},\tau)
             \cdot\nabla\psi^i({\bf x}, \tau)\right.
\nonumber\\
             + [v({\bf x}) - \mu]\, {\bar\psi}^i({\bf x},\tau)\,
                               \psi^i({\bf x},\tau)\biggr]
\nonumber\\
          +\ {1\over 2} \int d{\bf x}\,d{\bf y}\ u({\bf x}-{\bf y})\ 
             {\bar\psi}^i({\bf x},\tau)\,{\bar\psi}^j({\bf y},\tau)\,
             \psi^j({\bf y},\tau)\,
\nonumber\\
             \times\psi^i({\bf x},\tau)\quad.
\label{eq:2.2a}
\end{eqnarray}
Here $m$ is the electron mass, $\mu$ is the chemical potential, and
$u({\bf x}-{\bf y})$ is the electron-electron interaction potential.
We assume that the random potential $v({\bf x})$ is delta-correlated
and obeys a Gaussian distribution $P[v({\bf x)}]$ with second moment
\begin{equation}
\left\{v({\bf x})\,v({\bf y})\right\}_{dis} = {1\over 2\pi N_F \tau_{el}}\ 
                  \delta({\bf x} - {\bf y})\quad,
\label{eq:2.2b}
\end{equation}
where 
\begin{equation}
\{\ldots\}_{dis} = \int D[v]\ P[v]\ (\ldots)\quad,
\label{eq:2.2c}
\end{equation}
\end{mathletters}%
denotes the disorder average, $N_F$ is the bare density of states per spin 
at the Fermi level, and $\tau_{el}$ is the bare electron
elastic mean-free time. More realistic models to describe itinerant electron
magnetism including, e.g., band structure, can be considered along the same
lines. The salient points of our results, however, are due to long-wavelength
effects and hence do not depend on microscopic details like the band structure.
For our purposes it therefore is sufficient to study the model defined in
Eqs.\ (\ref{eqs:2.2}).

For describing magnetism, it is convenient and standard practice to
break the interaction part of the action $S$, which we denote by $S_{int}$,
into spin-singlet and spin triplet contributions, $S_{int}^{\,(s,t)}$. For
simplicity, we assume that the interactions are short-ranged in both of
these channels. In a metallic system this is justified due to screening,
and an effective model with a short-ranged interaction in both $S_{int}^{\,(s)}$
and $S_{int}^{\,(t)}$ can be derived starting from a bare Coulomb 
interaction.\cite{AGD} In order for this assumption to remain valid, our
discussion applies only to cases in which the disorder is weak enough
for the system to remain far from any metal-insulator transition that
might be present in the phase diagram and would lead to a breakdown of
screening.\cite{Mott} The spin-triplet interaction
$S_{int}^{\,(t)}$ describes interactions between spin density fluctuations. This
is what causes ferromagnetism, and it therefore makes sense to consider
this part of the action separately. We thus write
\begin{equation}
S = S_0 + S_{int}^{\,(t)}\quad,
\label{eq:2.3}
\end{equation}
with
\begin{mathletters}
\label{eqs:2.4}
\begin{equation}
S_{int}^{\,(t)} = {\Gamma_t\over 2} \int d{\bf x}\,d\tau\ {\bf n}_s({\bf x},\tau)
                 \cdot {\bf n}_s({\bf x},\tau)\quad,
\label{eq:2.4a}
\end{equation}
where ${\bf n}_s$ is the electron spin density vector with components,
\begin{equation}
n_s^a({\bf x},\tau) = {1\over 2}\ {\bar\psi}^i({\bf x},\tau)\,
     \sigma^a_{ij}\,\psi^j({\bf x},\tau)\quad.
\label{eq:2.4b}
\end{equation}
Here the $\sigma^a$ are the Pauli matrices, and $\Gamma_t$ is the
spin triplet interaction amplitude that is related to the interaction
potential $u$ in Eq.\ (\ref{eq:2.2a}) via
\begin{equation}
\Gamma_t = \int d{\bf x}\ u({\bf x}) \quad.
\label{eq:2.4c}
\end{equation}
\end{mathletters}%
For simplicity we have assumed a
point-like interaction so that $\Gamma_t$ is simply a number. A generalization
to a more realistic short-range interaction would be straightforward.
$S_0$ in Eq.\ (\ref{eq:2.3}) contains all other contributions to the
action. It reads explicitly, 
\begin{mathletters}
\label{eqs:2.5}
\begin{eqnarray}
S_0 = \int_0^{\beta} d\tau \int d{\bf x}\ \biggl[
   {\bar\psi}^i({\bf x},\tau)\,{\partial\over\partial\tau}\,\psi^i({\bf x},\tau)\quad\quad
                                                               \quad\quad
\nonumber\\
   - {1\over 2m}\,\nabla {\bar\psi}^i({\bf x},\tau)
             \cdot\nabla\psi^i({\bf x}, \tau)
\nonumber\\
   - [v({\bf x}) - \mu]\, {\bar\psi}^i({\bf x},\tau)\,
                            \psi^i({\bf x},\tau)\biggr]
\nonumber\\
   - {\Gamma_s\over 2} \int_0^{\beta} d\tau \int d{\bf x}\ 
         n_c({\bf x},\tau)\,n_c({\bf x},\tau)\quad,
\label{eq:2.5a}
\end{eqnarray}
with $n_c$ the electron charge or number density,
\begin{equation}
n_c({\bf x},\tau) = {\bar\psi}^i({\bf x},\tau)\, \psi^i({\bf x},\tau)\quad,
\label{eq:2.5b}
\end{equation}
\end{mathletters}%
and $\Gamma_s$ the spin-singlet interaction amplitude.

\subsection{Order parameter field theory}
\label{subsec:II.B}

Continuous {\it thermal} phase transitions are usually described by deriving
an LGW theory, i.e. an effective 
field theory for the long wavelength order parameter fluctuations
or critical modes.\cite{WilsonKogut} The physical idea
behind this approach is that these fluctuations, which are slowly varying 
in space, determine the behavior near the critical point. The same
philosophy has been applied to quantum phase transitions, with the only
difference being that the critical modes are now slowly varying in both
space and time. We will use
this approach here, motivated in part by previous work on clean itinerant
electronic systems.\cite{Hertz} We mention, however, that in general one
should worry about both the critical modes, and all other slow or soft 
modes, even if these other soft modes are not `critical' in the sense
that they change their character at the phase transition.
While this concern is not confined to quantum phase transitions, we will
argue below that for these it poses a more serious problem than for
thermal phase transitions since at $T=0$ there are more soft modes than
at finite temperature.
We will see that in the present problem such additional modes are 
indeed present and
lead to complications within the framework of an LGW theory. For the
problem under consideration, however, these complications can be overcome.

The techniques for deriving an order parameter field theory, starting
with Eqs.\ (\ref{eq:2.3}) through (\ref{eqs:2.5}), are standard.\cite{Hertz} 
We decouple the four fermion term in $S_{int}^{\,(t)}$ by introducing a
classical vector field ${\bf M}({\bf x},\tau)$ whose average is
proportional to the magnetization $m$, and performing a Hubbard-Stratonovich
transformation. All degrees of freedom other than ${\bf M}$ are then
integrated out. This procedure in particular integrates out the soft
diffusive modes or `diffusons' that are inherent to a disordered fermion 
system.\cite{LeeRama,R} These are the additional soft modes mentioned
above. We obtain the partition function $Z$ in the form,
\begin{mathletters}
\label{eqs:2.6}
\begin{equation}
Z = e^{-F_0/T} \int D[{\bf M}]\ \exp\left(-\Phi[{\bf M}]\right)\quad,
\label{eq:2.6a}
\end{equation}
with $F_0$ the noncritical part of the free energy. The LGW functional
$\Phi$ reads,
\begin{eqnarray}
\Phi[&&{\bf M}] = {\Gamma_t\over 2} \int dx\ {\bf M}(x)\cdot {\bf M}(x)
                \quad\quad\quad\quad\quad\quad
\nonumber\\
              &&- \ln \left<\exp\left[-\Gamma_t \int dx\ {\bf M}(x)\cdot
                              {\bf n}_s(x)\right]\right>_{S_0}\quad,
\label{eq:2.6b}
\end{eqnarray}
\end{mathletters}%
where we have adopted a four-vector notation with
$x = ({\bf x},\tau)$, and $\int dx = \int d{\bf x} \int_0^{\beta} d\tau$.
Here $\langle\ldots\rangle_{S_0}$ denotes an average taken with the
action $S_0$. A formal expansion of $\Phi$ in powers of ${\bf M}$ takes
the form,
\begin{mathletters}
\label{eqs:2.7}
\begin{eqnarray}
\Phi[{\bf M}] = {1\over 2} \int dx_1\,dx_2\ M_a(x_1)\,
                 \biggl[{\delta_{ab}\over \Gamma_t}\,\delta(x_1-x_2)
\nonumber\\
                            - \chi_{ab}^{\,(2)}(x_1,x_2)\biggr]\,M_b(x_2)
\nonumber\\
  +\ {1\over 3!} \int dx_1\,dx_2\,dx_3\ \chi_{abc}^{\,(3)}(x_1,x_2,x_3)\ 
\nonumber\\
                           \times M_a(x_1)\,M_b(x_2)\,M_c(x_3)
\nonumber\\
  -\ {1\over 4!} \int dx_1\,dx_2\,dx_3\,dx_4\ 
              \chi_{abcd}^{\,(4)}(x_1,x_2,x_3,x_4)
\nonumber\\
                     \times M_a(x_1)\,M_b(x_2)\,M_c(x_3)\,M_d(x_4)
\nonumber\\
  +\ O(M^5)\quad,
\label{eq:2.7a}
\end{eqnarray}
where we have scaled $M$ with $\Gamma_t^{-1}$. The coefficients $\chi^{(l)}$
in Eq.\ (\ref{eq:2.7a}) are correlation functions for a system with a
particular realization of the disorder (i.e., they are not translationally
invariant). They are defined as,
\begin{equation}
\chi_{a_1\ldots a_l}^{\,(l)}(x_1,\ldots,x_l) = \langle n_s^{a_1}(x_1)\cdots
                       n_s^{a_l}(x_l)\rangle^c_{S_0}\quad,
\label{eq:2.7b}
\end{equation}
\end{mathletters}%
where the superscript `c' denotes a cumulant or connected correlation
function. For our simple model, the reference ensemble with action $S_0$,
whose correlation functions are the $\chi^{(l)}$, consists of free
electrons with disorder and a short-ranged spin-singlet model 
interaction. As mentioned
in Sec.\ \ref{subsec:II.A} above, the model can be made more realistic, if
desired, by e.g. including a realistic band structure. In order to do so,
one would simply replace the $\chi^{(l)}$ above with correlation functions
for band electrons.

Equations\ (\ref{eqs:2.6}) and (\ref{eqs:2.7}) define an order parameter
field theory for the paramagnet-to-ferromagnet phase transition in a
disordered itinerant electronic system. The disorder has {\it not} been
averaged over yet, and so the $n$-point correlation functions in Eqs.\ 
(\ref{eqs:2.7}) depend explicitly on the particular realization of the
randomness in the system. To proceed, we now formally carry out the
disorder average of the free energy, using replicas and a cumulant
expansion, and keeping terms up to $O(M^4)$ in the replicated LGW
functional.\cite{Grinstein}
With $\alpha,\beta,\ldots=1,\ldots,n$ ($n\rightarrow 0$) the replica labels,
we obtain the following LGW functional for the $\alpha^{th}$ copy of
the replicated system,
\begin{mathletters}
\label{eqs:2.8}
\begin{eqnarray}
\Phi^{\alpha}[{\bf M}] = \sum_{l=2}^{\infty}\Phi^{\alpha}_l[{\bf M}]
            \quad\quad\quad\quad\quad\quad\quad\quad\quad\quad\quad\quad
\nonumber\\
  = {1\over 2} \int dx_1\,dx_2\ X^{(2)}_{ab}(x_1,x_2)\,M_a^{\alpha}(x_1)\,
                                                           M_b^{\alpha}(x_2)
\nonumber\\
  +\ {1\over 3!} \int dx_1\,dx_2\,dx_3\ X^{(3)}_{abc}(x_1,x_2,x_3)
              \quad\quad\quad
\nonumber\\
             \times M_a^{\alpha}(x_1)\,M_b^{\alpha}(x_2)\,M_c^{\alpha}(x_3)
\nonumber\\
  -\ {1\over 4!} \int dx_1\,dx_2\,dx_3\,dx_4\ 
               X^{(4)\,\alpha\beta}_{abcd}(x_1,x_2,x_3,x_4)
\nonumber\\
 \times M_a^{\alpha}(x_1)\,M_b^{\alpha}(x_2)\,M_c^{\beta}(x_3)\,M_d^{\beta}(x_4)
\nonumber\\
  + O(M^5)\quad,
\label{eq:2.8a}
\end{eqnarray}
where $\Phi^{\alpha}_l$ denotes the contribution of order $M^l$, and the
coefficients are given in terms of disorder averaged correlation functions,
\begin{equation}
X^{(2)}_{ab}(x_1,x_2) = {\delta_{ab}\over \Gamma_t}\,\delta(x_1-x_2) -
                          \{\chi_{ab}^{\,(2)}(x_1,x_2)\}_{dis}\quad,
\label{eq:2.8b}
\end{equation}
\begin{equation}
X^{(3)}_{abc}(x_1,x_2,x_3) = \{\chi_{abc}^{\,(3)}(x_1,x_2,x_3)\}_{dis}\quad,
\label{eq:2.8c}
\end{equation}
\begin{eqnarray}
X^{(4)\,\alpha\beta}_{abcd}(x_1,&&x_2,x_3,x_4) = 
   \delta_{\alpha\beta}\{\chi_{abcd}^{\,(4)}(x_1,x_2,x_3,x_4)\}_{dis}
\nonumber\\
  &&+\ 3\{\chi_{ab}^{\,(2)}(x_1,x_2)\,\chi_{cd}^{\,(2)}(x_3,x_4)\}_{dis}^c
                                                              \quad.
\label{eq:2.8d}
\end{eqnarray}
\end{mathletters}%
Again, the superscript `c' means that only the connected part,
this time with respect
to the disorder average, of the correlation functions $\{\ldots\}_{dis}^c$
should be considered. Notice that we have separated the quantum mechanical
and disorder averages, and therefore must deal with two different kinds
of cumulants: In Eqs.\ (\ref{eqs:2.7}), we have $\langle ab\rangle^c =
\langle ab\rangle - \langle a\rangle\,\langle b\rangle$, etc. while
in Eqs.\ (\ref{eqs:2.8}) we have 
$\{ab\}_{dis}^c = \{ab\}_{dis} - \{a\}_{dis}\{b\}_{dis}$, etc.

\subsection{The coefficients}
\label{subsec:II.C}

As it stands, Eq.\ (\ref{eq:2.8a}) is just a formal nonlocal field theory
that is not very useful. Normally, one would proceed by localizing the
individual terms in Eq.\ (\ref{eq:2.8a}) about a single point in space and
time, and expanding the correlation functions in powers of gradients. Due
to what are often called weak localization effects,\cite{WeakLoc} this
is not an option for the current case of a disordered metal, even if we
stay far away from any metal-insulator transition. Structurally it is
easy to show, using either many-body perturbation theory or field theoretic
methods, that the disorder averaged $n$-point correlation function,
$\{\chi^{(l)}\}_{dis}$, in
wavenumber space has a weak localization correction of a form that can
be written schematically as,
\begin{equation}
\delta\chi^{(l)}({\bf q},\Omega_n) \sim {1\over V}\sum_{k>q} T
      \sum_{\omega_n>\Omega_n}\ {1\over ({\bf k}^2 + \omega_n)^l}\quad,
\label{eq:2.9}
\end{equation}
with $V$ the system volume, $k=\vert{\bf k}\vert$, 
$\omega_n$ and $\Omega_n$ Matsubara frequencies, and $({\bf q},\Omega_n)$ an 
infrared cutoff. Although, strictly speaking, $\delta\chi^{(l)}$ depends on
$l-1$ external momenta and frequencies, we have schematically represented 
these by a single `typical' momentum-frequency $({\bf q},\Omega_n)$. Since
we will only be interested in the scale dimension of $\{\chi^{(l)}\}_{dis}$, 
this is sufficient. In terms of diagrams for the field theory reviewed in 
Ref.\ \onlinecite{R}, the dominant contribution to 
$\{\chi^{(l)}\}_{dis}(q,\Omega_n=0)$ for $q\rightarrow 0$ is shown
\begin{figure}
\epsfxsize=6.25cm
\epsfysize=4.8cm
\epsffile{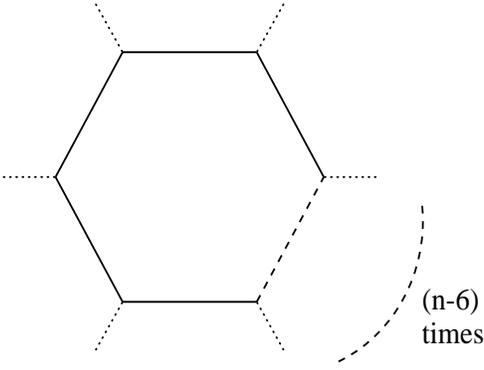}
\vskip 0.5cm
\caption{Diagrammatic structure of the leading IR singular contribution to
 the correlation function $\chi^{(l)}(q,\Omega_n=0)$. The
 straight lines denote the diffusive propagators of the field 
 theory reviewed in Ref.\ \protect\onlinecite{R},
 and the dotted lines denote external legs.}
\label{fig:1}
\end{figure}
in Fig.\ \ref{fig:1}. For what follows it is very important to notice that 
$\delta\chi^{(l)}(q,\Omega_n=0)$ is a nonanalytic function of $q$.
 
To illustrate this point in detail, let us consider the two-point correlation
function, $\chi_{ab}^{\,(2)}$, in Eq.\ (\ref{eq:2.8b}). Because of 
translational invariance on average in space and time, as well as rotational
invariance in space, we can write,
\begin{mathletters}
\label{eqs:2.10}
\begin{equation}
\{\chi_{ab}^{\,(2)}(x_1,x_2)\}_{dis} = \delta_{ab}\,\chi_s(\vert{\bf x}_1 -
      {\bf x}_2\vert,\tau_1 - \tau_2)\quad,
\label{eq:2.10a}
\end{equation}
and we define the Fourier transform
\begin{eqnarray}
\chi_s({\bf q},\Omega_n) = \int d(&&{\bf x}_1-{\bf x}_2)\,d(\tau_1-\tau_2)\ 
            e^{-i{\bf q}\cdot ({\bf x}_1-{\bf x}_2)}
\nonumber\\
                                     &&\times e^{i\Omega_n (\tau_1-\tau_2)} 
            \ \chi_s(x_1-x_2)\quad.
\label{eq:2.10b}
\end{eqnarray}
\end{mathletters}%
$\chi_s$ is the disorder averaged spin susceptibility of the reference
system whose action is given by $S_0$. Particle number conservation
implies that at small frequency and wavenumber it has a diffusive structure,
\begin{mathletters}
\label{eqs:2.11}
\begin{equation}
\chi_s({\bf q},\Omega_n) = \chi_0(q)\ {D\,q^2
                \over \vert\Omega_n\vert + D\,q^2}\quad,
\label{eq:2.11a}
\end{equation}
where $D$ is the spin diffusion coefficient of the reference ensemble, and
$\chi_0(q)$ is the static spin susceptibility. In a system
with a conserved order parameter, the frequency must be taken to zero
before the wavenumber in order to reach criticality, so in the critical
region we have $\vert\Omega_n\vert << D\,q^2$. Note that it
is our restriction to nonmagnetic disorder that ensures a conserved order
parameter. We will come back to this point below. In the critical limit,
we can thus expand,
\begin{equation}
\chi_s({\bf q},\Omega_n) = \chi_0(q)\ \left[ 1 -
      \vert\Omega_n\vert/Dq^2 + \ldots\right]\quad.
\label{eq:2.11b}
\end{equation}
\end{mathletters}%

The static spin susceptibility, $\chi_0(q)$, has been discussed in some
detail before.\cite{fatpaper} For any system with a nonvanishing
interaction amplitude in the spin-triplet channel, $\Gamma_t \neq 0$, there 
is a diffusive or weak localization correction to the bare 
susceptibility. Although our bare 
reference ensemble does not contain such an interaction amplitude, perturbation
theory will generate one as long as there is a nonvanishing interaction 
amplitude $\Gamma_s$ in the spin-singlet channel.\cite{fatpaper}
Effectively, we therefore need to use the spin
susceptibility for a system with a nonvanishing $\Gamma_t$ in 
Eq.\ (\ref{eq:2.11b}). The explicit calculation is given in Appendix\ 
\ref{app:A}, and the result is,
\begin{mathletters}
\label{eqs:2.12}
\begin{equation}
\chi_0(q\rightarrow 0) = c_0 - c_{d-2}\,q^{d-2} - c_2\,q^2 + \ldots\quad,
\label{eq:2.12a}
\end{equation}
where the $c_i$ are positive constants. The nontrivial, and for our purposes
most interesting, contribution in Eq.\ (\ref{eq:2.12a}) is the nonanalytic
term $\sim q^{d-2}$. Its existence implies that the standard gradient
expansion mentioned above Eq.\ (\ref{eq:2.9}) does not exist. The
physical interpretation of this term is that, effectively, there is a
long-range interaction between the order parameter fluctuations, which
in real space takes the form $r^{-2d+2}$. This is a
phenomenon that is special to zero temperature. At finite temperature,
when one has to perform a frequency sum rather than a frequency integral
to calculate the correlation funcion,
the nonanalytic term is replaced by a term of the schematic structure,
\begin{equation}
q^{d-2} \rightarrow (q^2 + T)^{(d-2)/2}\quad,
\label{eq:2.12b}
\end{equation}
\end{mathletters}%
so for $T>0$ an analytic expansion about $q=0$ exists, and the standard
local LGW functional is obtained.

Next we consider the cubic term in Eq.\ (\ref{eq:2.8a}). Rotational symmetry
in spin space allows us to write $\Phi^{\alpha}_3$ as,
\begin{mathletters}
\label{eqs:2.13}
\begin{eqnarray}
\Phi^{\alpha}_3 [{\bf M}] = {1\over 3!} \int dx_1\,dx_2\,dx_3\ 
                                                     u_3(x_1,x_2,x_3)\ 
\nonumber\\
\times{\bf M}^{\alpha}(x_1)\cdot\left({\bf M}^{\alpha}
        (x_2)\times{\bf M}^{\alpha}(x_3)\right)\quad,
\label{eq:2.13a}
\end{eqnarray}
with
\begin{equation}
u_3(x_1,x_2,x_3) = {1\over 6}\ \epsilon_{abc}\ \{\chi_{abc}^{(3)}
                       (x_1,x_2,x_3)\}_{dis}\quad,
\label{eq:2.13b}
\end{equation}
where $\epsilon_{abc}$ is the Levi-Civita tensor. According to 
Eq.\ (\ref{eq:2.9}), 
\begin{equation}
u_3(q\rightarrow 0,\Omega_n=0) = u_3^{(d-4)}\,q^{d-4} + u_3^{(0)}\quad,
\label{eq:2.13c}
\end{equation}
\end{mathletters}%
with $u_3^{(d-4)}$ and $u_3^{(0)}$ finite numbers.
Note that this term is divergent for $q\rightarrow 0$ for $d<4$.

The quartic term in Eq.\ (\ref{eq:2.8a}) is written as a sum of two terms,
\begin{mathletters}
\label{eqs:2.14}
\begin{equation}
\Phi^{\alpha}_4[{\bf M}] = \Phi^{\alpha\,(1)}_4[{\bf M}] +
                           \Phi^{\alpha\,(2)}_4[{\bf M}] \quad,
\label{eq:2.14a}
\end{equation}
with
\begin{eqnarray}
\Phi^{\alpha\,(1)}_4[{\bf M]} = -\,&&{1\over 24} \int dx_1\ldots dx_4\ 
          \{\chi_{abcd}^{\,(4)}(x_1,\ldots,x_4)\}_{dis}\ 
\nonumber\\
&&\times M_a^{\alpha}(x_1)\,M_b^{\alpha}(x_2)\,M_c^{\alpha}(x_3)\,
                                                     M_d^{\alpha}(x_4)\quad,
\nonumber\\
\label{eq:2.14b}
\end{eqnarray}
and
\begin{eqnarray}
\Phi^{\alpha\,(2)}_4[{\bf M}] = {1\over 8} \int dx_1\,dx_2\,dx_3\,dx_4\ 
\nonumber\\
     \times\{\chi_{ab}^{\,(2)}(x_1,x_2)\,\chi_{cd}^{\,(2)}(x_3,x_4)\}_{dis}^c
\nonumber\\
\times M_a^{\alpha}(x_1)\,M_b^{\alpha}(x_2)\,M_c^{\alpha}(x_3)\,M_d^{\alpha}(x_4)
                                                       \quad.
\label{eq:2.14c}
\end{eqnarray}
\end{mathletters}%
For our purposes we only need to know the degrees of divergence of the
coefficients in Eqs.\ (\ref{eqs:2.14}). We can therefore formally consider
the local, static limit of Eq.\ (\ref{eq:2.14b}) (even though it does not
necessarily exist), and write
\begin{mathletters}
\label{eqs:2.15}
\begin{eqnarray}
\Phi_4^{\alpha\,(1)}[{\bf M}] = {T\over 24V}\,u_4\sum_{{\bf q}_1,{\bf q}_2,
    {\bf q}_3}\ \sum_{n_1,n_2,n_3}\quad\qquad\qquad
\nonumber\\
\times\left({\bf M}^{\alpha}({\bf q}_1,\omega_{n_1})\cdot
    {\bf M}^{\alpha}({\bf q}_2,\omega_{n_2})\right)
\nonumber\\
\times \left({\bf M}^{\alpha}({\bf q}_3,\omega_{n_3})
  \cdot {\bf M}^{\alpha}(-{\bf q}_1 - {\bf q}_2 - {\bf q}_3,-\omega_{n_1}\right.
\nonumber\\
          \left. -\omega_{n_2} - \omega_{n_3})\right)\quad,
\label{eq:2.15a}
\end{eqnarray}
where
\begin{equation}
M_a^{\alpha}({\bf q},\omega_n) = {1\over\sqrt{\beta V}} \int d{\bf x}\,
    d\tau\ e^{-i{\bf q}\cdot{\bf x} + i\omega_n\tau}\,M_a^{\alpha}
    ({\bf x},\tau)\quad,
\label{eq:2.15b}
\end{equation}
and
\begin{equation}
u_4 = -\lim_{\{{\bf q}\}\rightarrow 0} \{\chi_{aaaa}^{\,(4)}\}_{dis}
       ({\bf q}_1,{\bf q}_2,{\bf q}_3; \omega_{n_1}\!=\omega_{n_2}\!=
                                        \omega_{n_3}\!=0).
\label{eq:2.15c}
\end{equation}
According to Eq.\ (\ref{eq:2.9}),
\begin{equation}
u_4(q\rightarrow 0,\Omega_n=0) = u_4^{(d-6)}\,q^{d-6} + u_4^{(0)}\quad,
\label{eq:2.15d}
\end{equation}
\end{mathletters}%
so $u_4$ diverges in this limit for $d < 6$, and is finite for $d>6$.
With a more accurate representation for $u_4$ than Eq.\ (\ref{eq:2.9}) one
also finds a term $\sim q^{d-4}$, but this will be of no relevance for
what follows.

We now consider $\Phi_4^{\alpha\,(2)}$ given by Eq.\ (\ref{eq:2.14c}). Its
most interesting feature is its frequency structure that follows from
$\{\chi_{ab}^{\,(2)}(x_1,x_2)\,\chi_{cd}^{\,(2)}(x_3,x_4)\}_{dis}^c$ being
a function of $\tau_1-\tau_2$ and $\tau_3-\tau_4$.\cite{connfootnote} 
This implies that
there are effectively two free $\tau$ integrals in $\Phi_4^{\alpha\,(2)}$,
unlike the case of $\Phi_4^{\alpha\,(1)}$, where there is only one. In
frequency space, $\Phi_4^{\alpha\,(2)}$ can be written,
\begin{eqnarray}
\Phi_4^{\alpha\,(2)}[{\bf M}] = -{1\over 8} \int d{\bf x}_1\,d{\bf x}_2\,d{\bf x}_3\,d{\bf x}_4\qquad\qquad\qquad
\nonumber\\
\sum_{n_1 n_2}\ \{\chi_{ab}^{\,(2)}({\bf x}_1,{\bf x}_2;\omega_{n_1})\,
               \chi_{cd}^{\,(2)}({\bf x}_3,{\bf x}_4;\omega_{n_2})\}_{dis}^c
\nonumber\\
               \ M_a^{\alpha}({\bf x}_1,\omega_{n_1})\,
                M_b^{\alpha}({\bf x}_2,-\omega_{n_1})\,
M_c^{\alpha}({\bf x}_3,\omega_{n_2})\,
\nonumber\\
\times M_d^{\alpha}({\bf x}_4,-\omega_{n_2})\quad.
\label{eq:2.16}
\end{eqnarray}
We see that $\Phi_4^{\alpha\,(1)}$ carries an extra factor of $T$ compared to
$\Phi_4^{\alpha\,(2)}$. The correlation function in 
Eq.\ (\ref{eq:2.16}) has been calculated for noninteracting
disordered electrons.\cite{JAS} The important result of these authors
was that the Fourier transform of the correlation function at 
$\omega_{n_1} = \omega_{n_2} = 0$ is finite. We have 
convinced ourselves that including
interactions does not change this result. $\Phi_4^{\alpha\,(2)}[{\bf M}]$
can then be replaced by
\begin{eqnarray}
\Phi_4^{\alpha\,(2)}[{\bf M}] = -{v_4\over 8} \int d{\bf x} \sum_{n_1 n_2}
      \vert{\bf M}^{\alpha}({\bf x},\omega_{n_1})\vert^2\,
\nonumber\\
      \times\vert{\bf M}^{\beta}({\bf x},\omega_{n_2})\vert^2\quad,
\label{eq:2.17}
\end{eqnarray}
with $v_4$ a finite coefficient. The physical meaning of $\Phi_4^{\alpha\,(2)}$
is easily determined. Consider the local in space and time contribution to
the disorder average of the coefficient $X^{(2)}$ in Eq.\ (\ref{eq:2.8a}),
i.e. the term $t_0 = 1/\Gamma_t - \chi_0(q=0)$, which determines the 
distance from the critical point in the Gaussian theory. If we make $t_0$
a random function of space with a Gaussian distribution and integrate out
that randomness, then we obtain a term with the structure of 
$\Phi_4^{\alpha\,(2)}$. This term in our action thus represents a
`random mass' term, reflecting the fluctuations in the location of the
critical point mentioned in the introduction.

We conclude this section with a discussion of why the diagram shown in
Fig.\ \ref{fig:1} gives the most important contribution to the correlation
function $\{\chi^{(l)}\}_{dis}$. 
The field theory of Ref.\ \onlinecite{R} allows
for a systematic loop expansion for these correlation functions, with
the diagram shown in Fig.\ \ref{fig:1} the one-loop contribution.
If we consider higher-loop corrections to this, we need to distinguish
between skeleton diagrams and insertions. Insertions will produce finite
(in $d\geq 2$) renormalizations of the one-loop result. In the skeletons,
each additional loop adds one frequency-momentum integral, and one
independent propagator. Since the propagators are at most a diffusion
pole squared,\cite{R} this means that each loop adds effectively a
factor of $q^{d-2}$ to $\{\chi^{(l)}\}_{dis}$, 
and hence we need to consider the
lowest loop contribution for a given type of term. For the static terms,
i.e. those of zeroth order in the frequency $\Omega_n$, the lowest
nonvanishing contribution is at one-loop order since the zero-loop
terms are of $O(\Omega_n)$. For instance, the zero-loop contribution
to $\{\chi^{(2)}\}_{dis}$ 
is $\sim\vert\Omega_n\vert/(\vert\Omega_n\vert + Dq^2)$,
while the one-loop contribution adds the static piece to yield
Eq.\ (\ref{eq:2.11a}). We thus conclude that the most infrared 
divergent contribution to the
static $l$-point correlation function is given by the diagram shown in
Fig.\ \ref{fig:1} with dressed internal propagators, and for just getting
the power of the leading divergence it suffices to use bare internal
propagators.

In the critical limit the zero-loop
contribution to $\{\chi^{(2)}\}_{dis}$ is proportional to $\Omega_n/q^2$, see
Eq.\ (\ref{eq:2.11b}). In the next section we will discuss a fixed point
where the dynamical exponent is $z=d$, i.e. $\Omega_n$ scales like $q^d$
at criticality. In the critical limit, the term of $O(\Omega_n)$ therefore
shows the same scaling behavior as the term of $O(\Omega_n^0)$, 
Eq.\ (\ref{eq:2.12a}). The same is true for all higher correlation functions.
It is easy to show that the zero-loop contribution to $\{\chi^{(l)}\}_{dis}$ 
goes
like $\Omega_n/q^{2l-2}$, while the one-loop contribution goes like
$q^{d-(2l-2)}$. Anticipating again that $\Omega_n$ scales like $q^d$ at
criticality, these two contributions have the same scaling behavior for
all values of $l$. All higher loop contributions will be less important,
as we have seen above. Also, higher orders in an expansion in powers of
$\Omega$ can be neglected, since $\Omega_n\sim q^2$ in the diffusive
propagators. Hence, each additional power of $\Omega_n$ will lead to
a factor that scales like $q^{d-2}$ at criticality and is less relevant 
than the terms of $O(\Omega_n^0)$ and $O(\Omega_n)$. We thus conclude
that the most relevant term of $O(M^l)$ in the GLW functional has a
coefficient $u_l$ that behaves effectively like
\begin{equation}
u_l \sim q^{d-2(l-1)}\quad.
\label{eq:2.18}
\end{equation}
This covers both the terms of $O(\Omega_n^0)$ and $O(\Omega_n)$, and
all higher powers of $\Omega_n$ are less important. The leading
term for $q\rightarrow 0$, shown in Eq.\ (\ref{eq:2.18}), 
comes from the contribution of the $l$-th moment to the cumulant $X^{(l)}$, 
i.e. from $\{\chi^{(l)}\}_{dis}$, while the other contributions,
i.e. the subtraction terms in a given cumulant, are less divergent.
However, due to the frequency structure of these terms that was explained 
above using the example of $X^{(4)}$, it is not obvious that they are
unimportant for the critical behavior. We will analyze this point in the
next section, and will find that only the subtraction term in the quartic
cumulant, which is given explicitly in Eq.\ (\ref{eq:2.17}), is important
in that respect and thus needs to be kept.
 
\section{The Critical Behavior}
\label{sec:III}

In the first part of this section we discuss the critical behavior of the
Gaussian part of the theory defined by 
Eqs.\ (\ref{eqs:2.8}) - (\ref{eq:2.17}). The
renormalization group properties of the Gaussian fixed point are also
discussed. We then analyze the non-Gaussian terms in the field theory, and
show that they are irrelevant, in the renormalization group sense, with
respect to the Gaussian fixed point for all
dimensions $d>2$, except for a marginal operator in $d=4$. 
This implies that the Gaussian theory yields the exact
critical behavior for all of these dimensions, except for logarithmic
corrections to scaling in $d=4$ and $d=6$ 
that are discussed in Appendix\ \ref{app:B}. We then construct the equation
of state near the critical point, which requires a more detailed knowledge
of the non-Gaussian terms in the field theory since it is determined
in part by dangerous irrelevant variables. We conclude this section with a
discussion of the specific heat near the quantum critical point, and with
a discussion of the quantum-to-classical crossover behavior.

\subsection{The Gaussian fixed point}
\label{subsec:III.A}

According to Eqs.\ (\ref{eqs:2.8}) - (\ref{eqs:2.12}), the Gaussian part
of $\Phi^{\alpha}$ is,
\begin{mathletters}
\label{eqs:3.1}
\begin{eqnarray}
\Phi^{\alpha}_2[{\bf M}] &&= {1\over 2} \sum_{\bf q}\sum_{\omega_n}
     {\bf M}^{\alpha}({\bf q},\omega_n)\,\left[t_0 + a_{d-2}\,q^{d-2} \right.
\nonumber\\
      &&\left. +\ a_2\,q^2 + a_{\omega}\,\vert\omega_n\vert/q^2\right]\,
     \cdot {\bf M}^{\alpha}(-{\bf q},-\omega_n)\ ,
\label{eq:3.1a}
\end{eqnarray}
where 
\begin{equation}
t_0 = 1 - \Gamma_t\,\chi_s({\bf q}\rightarrow 0,\omega_n=0)\quad,
\label{eq:3.1b}
\end{equation}
\end{mathletters}%
is the bare distance from the critical point, and $a_{d-2}$, $a_2$, and
$a_{\omega}$ are positive constants.

We first analyze the critical behavior implied by Eqs.\ (\ref{eqs:3.1}).
Later we will show that for $d>2$ fluctuations are irrelevant, and the
critical behavior found this way is exact for these dimensions. By
inspection of the Gaussian LGW functional in Eq.\ (\ref{eq:3.1a}) one
obtains,
\begin{mathletters}
\label{eqs:3.2}
\begin{equation}
\nu=\cases{1/(d-2)& for $2<d<4$\cr%
               1/2& for $d>4$\cr}%
     \quad,
\label{eq:3.2a}
\end{equation}
\begin{equation}
\eta = \cases{4-d& for $2<d<4$\cr%
                0&   for $d>4$\cr}%
        \quad,
\label{eq:3.2b}
\end{equation}
\begin{equation}
z = \cases{d& for $2<d<4$\cr%
           4& for $d>4$\cr}%
        \quad.
\label{eq:3.2c}
\end{equation}
\end{mathletters}%
Here $\nu$ is the correlation length exponent, defined by 
$\xi\sim t^{-\nu}$, with $t$ the dimensionless distance from the
critical point. $\eta$ is the exponent that determines the wavenumber
dependence of the order parameter susceptibility at criticality,
$\langle M_a({\bf q},0)\,M_a({-\bf q},0)\rangle \sim q^{-2+\eta}$. $z$ is the
dynamical scaling exponent that characterizes critical slowing down by
relating the divergence of the relaxation time, $\tau_r$, to that of the
correlation length, $\tau_r \sim \xi^z$.

Let us discuss, for later reference, the critical behavior given by
Eqs.\ (\ref{eqs:3.1}) and (\ref{eqs:3.2}) from a renormalization group
point of view. Let $b$ be the renormalization group length rescaling
factor. Under renormalization, all quantities change according to
$A\rightarrow A(b)= b^{[A]}\,A$, with $[A]$ the scale dimension of $A$. The
scale dimension of the order parameter is,
\begin{mathletters}
\label{eqs:3.3}
\begin{equation}
[{\bf M}({\bf q},\omega_n)] = -1+\eta/2 \quad,
\label{eq:3.3a}
\end{equation}
or, equivalently,
\begin{equation}
[{\bf M}({\bf x},\tau)] = (d+2)/2 \quad.
\label{eq:3.3b}
\end{equation}
\end{mathletters}%
At the critical fixed point, $a_{\omega}$ and either $a_{d-2}$ (for
$2<d<4$), or $a_2$ (for $d>4$) are not renormalized, i.e. there scale
dimensions are zero. Using this, and $[q]=1$, $[\omega_n]=z$
immediately yields Eqs.\ (\ref{eq:3.2b},\ \ref{eq:3.2c}). 
Equation\ (\ref{eq:3.2a}) follows from the relevance of $t_0$, or its
renormalized counterpart, $t$,
at the critical fixed point. That is, the scale dimension of $t$ is
positive and given by $1/\nu\equiv [t] = 2-\eta = d-2$.

\subsection{The non-Gaussian terms}
\label{subsec:III.B}

We now show that all of the non-Gaussian terms in the field theory are
renormalization group irrelevant with respect to the Gaussian fixed point
discussed in the last subsection. Let us first consider the term of order
$M^3$. From Eq.\ (\ref{eq:2.13c}) we see that the most relevant coefficient
at that order is $u_3^{(d-4)}$ for $d<4$, and $u_3^{(0)}$ for $d>4$.
For simplicity, we will use
the symbol $u_3$ for the most relevant coefficient in a given dimension,
i.e. $u_3$ denotes $u_3^{(d-4)}$ for $d<4$, and $u_3^{(0)}$ for $d>4$.
For $d<4$ we then have to assign a scale dimension to the cutoff wavenumber
$q$. The most obvious choice is to identify $q$ with the inverse correlation
length, $\xi^{-1}$, which makes $[q]=1$. While we will see later that this
identification is not correct for all values of $d$, it will turn out that
it provides an upper limit for $[q]$. Assuming, then, $[q]\leq 1$, which
will be justified in Sec.\ \ref{subsec:III.C} below,
and using Eq.\ (\ref{eqs:3.3}), we see that the scale dimension of the
effective coefficient of the term of order $M^3$ is bounded by,
\begin{equation}
[u_3] \leq -(d-2)/2\quad.
\label{eq:3.4}
\end{equation}
Similarly, using Eqs.\ (\ref{eq:2.15d}) and (\ref{eqs:3.3}), one sees that
the coefficient $u_4$ in Eq.\ (\ref{eq:2.15a})
has a scale dimension bounded by,
\begin{mathletters}
\label{eqs:3.5}
\begin{equation}
[u_4] \leq \cases{-(d-2)& for $2<d<4$\cr%
                 -2(d-3)& for $4<d<6$\cr%
                  -d    & for $d>6$\cr}%
         \quad,
\label{eq:3.5a}
\end{equation}
and that the scale dimension of $v_4$ in Eq.\ (\ref{eq:2.17}) is,
\begin{equation}
[v_4] = -\vert 4-d\vert\quad.
\label{eq:3.5b}
\end{equation}
Here again we denote the most relevant coefficient in Eq.\ (\ref{eq:2.15d})
by $u_4$ for simplicity.
Equations\ (\ref{eq:3.4}) and (\ref{eqs:3.5}) imply that the cubic and
quartic terms in the field theory are renormalization group irrelevant
with respect to the Gaussian fixed point for all $d>2$ except for
$d=4$, where the coefficient $v_4$ is a marginal operator. As we explained
after Eq.\ (\ref{eq:2.17}), $v_4$ reflects the `random mass' contribution
to the action of the disordered magnet. We stress that this operator,
while being strongly relevant with respect to the {\it clean} Gaussian
fixed point, is not relevant with respect to the Gaussian fixed point of
the present LGW functional. 

It is an easy matter to analyze the behavior of the higher order terms
in the Landau expansion. The coefficient of the general term of order
$M^l$ in the LGW functional has a scale dimension bounded by,
\begin{equation}
[u_l] \leq \cases{-(d-2)(l-2)/2& for $2<d<4$\cr%
              -(l-2) - l(d-4)/2& for $4<d<6$\cr%
              -(l-4) - d(l-2)/2& for $d>6$\cr}%
        \quad.
\label{eq:3.5c}
\end{equation}
\end{mathletters}%
This holds for the pure moment contribution to the $l$-th cumulant, and we see
that all of the $u_l$ are irrelevant for $d>2$. For the coefficients of the
subtraction terms in the cumulants one easily convinces oneself that,
while their scale dimension initially increases with $d$ increasing
from $d=2$ (as does the scale dimension of $v_4$, see Eq.\ (\ref{eq:3.5b})),
it stays negative for all $d>2$. All of these terms are therefore
irrelevant operators as well. We conclude
that for $d>2$ there is a critical Gaussian fixed point
corresponding to a phase transition with Gaussian exponents (except
for logarithmic corrections to the Gaussian critical behavior in $d=4$
due to the marginal operator $v$, and, as we will see later, also in $d=6$), 
while for $d\leq 2$ the non-Gaussian
terms are potentially relevant. We will see in Sec.\ \ref{subsec:III.C}
below that the inequality in Eq.\ (\ref{eq:3.5c}) is actually an equality
for $2<d<4$, so that $d=2$ is the upper critical dimension, and in $d=2$
an {\it infinite} number of operators become marginal.

\subsection{The magnetization, and the magnetic susceptibility}
\label{subsec:III.C}

Although $u_4$ is an irrelevant operator for $d>2$, it is dangerously
irrelevant\cite{MEF} for the magnetization, since it determines the critical
behavior of the average magnetization, $m$, as a function of $t$ and 
an external magnetic field $H$. The technical reason for this is that
$m$ is a singular function of $u$ for $u\rightarrow 0$. Schematically,
the equation of state in mean-field theory is of the form,
\begin{equation}
t\,m + u_4\,m^3 = H\quad,
\label{eq:3.6}
\end{equation}
where we have suppressed all numerical prefactors. According to
Eq.\ (\ref{eq:2.15d}), $u_4$ diverges for $d<6$ as $q^{d-6}$. Furthermore,
if we were to keep higher terms of order $m^l$ in the equation of 
state, their coefficients would diverge as $q^{d-2l}$.
This implies that the cutoff $q$ scales like $m^{1/2}$, and effectively 
Eq.\ (\ref{eq:3.6}) reads,
\begin{equation}
t\,m + {\bar u}\,m^{d/2} + u_4^{(0)}\,m^3 = H\quad,
\label{eq:3.7}
\end{equation}
with $u_4^{(0)}$ from Eq.\ (\ref{eq:2.15d}), and
$\bar u$ another finite coefficient. From Eq.\ (\ref{eq:3.7})
we immediately obtain the exponents $\beta$ and $\delta$, defined as
$m(t,H=0)\sim t^{\beta}$, $m(t=0,H)\sim H^{1/\delta}$,
\begin{mathletters}
\label{eqs:3.8}
\begin{equation}
\beta = \cases{2/(d-2)& for $2<d<6$\cr%
                   1/2& for $d>6$\cr}%
     \quad,
\label{eq:3.8a}
\end{equation}
\begin{equation}
\delta = \cases{d/2& for $2<d<6$\cr%
                  3& for $d>6$\cr}%
     \quad.
\label{eq:3.8b}
\end{equation}
\end{mathletters}%
In $d=6$ logarithmic corrections to scaling occur, see Appendix \ref{app:B}.

Above we have used a general scaling argument to obtain Eq.\ (\ref{eq:3.7}).
For small disorder, the same result can also be derived 
explicitly by means of an infinite
resummation. As mentioned after Eq.\ (\ref{eq:3.6}), the term of $O(m^3)$ 
with its divergent coefficient $u_4\sim q^{d-6}$ in Eq.\ (\ref{eq:3.6}) is 
only the first in an infinite series of terms that behave like
$q^{d-2l}\,m^l$. Calculating the prefactors of the divergent coefficients,
one realizes that the divergencies are the consequence of an illegal
expansion of an equation of state of the form
\begin{equation}
t\,m + {1\over V}\sum_{\bf k}\,T\sum_{\omega_n} {{\rm const}\times m^3\over 
                          \left[(\omega_n + k^2)^2 + m^2\right]^2} = H\quad.
\eqnum{3.6'}
\label{eq:3.6'}
\end{equation}
Performing the integral one recovers Eq.\ (\ref{eq:3.7}).

Next we determine the functional form of the equation of state at
nonzero temperature in order to obtain a scaling equation for $m$ as
a function of $t$, $H$, and $T$. This can most easily be done by
utilizing Eq.\ (\ref{eq:2.12b}). From Eq.\ (\ref{eq:2.12a}) it follows
that the $tm$-term in the equation of state has a correction of the
form $m\,(q^2 + T)^{(d-2)/2} \sim m\,(m+T)^{(d-2)/2}$, where we have used
$q^2\sim m$ as explained after Eq.\ (\ref{eq:3.6}). Similarly, the
term $u\,m^3$ in Eq.\ (\ref{eq:3.6}) with $u\sim q^{d-6}$ gets replaced
by $m^3\,(m+T)^{(d-6)/2}$. At $T=0$ we recover Eq.\ (\ref{eq:3.7}). For
$T<m$ in suitable units, there are corrections of $O(T/m)$ to the
term $\sim m^{d/2}$ in that equation, while for $T>m$, $t$ gets replaced
by $t+T^{1/2\nu}$. All of these observations can be summarized in the
following homogeneity law,
\begin{mathletters}
\label{eqs:3.9}
\begin{equation}
m(t,T,H) = b^{-\beta/\nu}\,m(tb^{1/\nu}, Tb^{\phi/\nu}, Hb^{\delta\beta/\nu})\quad,
\label{eq:3.9a}
\end{equation}
with
\begin{equation}
\phi = 2\nu\quad.
\label{eq:3.9b}
\end{equation}
\end{mathletters}%
Similarly, the magnetic susceptibility, $\chi_m$, satisfies a 
homogeneity law,
\begin{mathletters}
\label{eqs:3.10}
\begin{equation}
\chi_m(t,T,H) = b^{\gamma/\nu}\,\chi_m(tb^{1/\nu}, Tb^{\phi/\nu}, Hb^{\delta\beta/\nu})\quad,
\label{eq:3.10a}
\end{equation}
with 
\begin{equation}
\gamma = \beta (\delta -1) = 1\quad.
\label{eq:3.10b}
\end{equation}

The Eqs.\ (\ref{eqs:3.9}) and (\ref{eqs:3.10}) warrant some discussion.
The scaling of $T$ in these equations follows directly from 
Eq.\ (\ref{eq:2.12b}). The effective scale dimension of $T$ in $m$
and $\chi_m$ is therefore $2$ and {\it not} $z$. The salient point is that
$z$ is determined by the scaling of $\Omega_n$ or $T$ with $q$ in the 
Gaussian action, and hence in the critical propagator. However, the 
magnetization is calculated at $\Omega_n=q=0$, and its leading temperature
dependence is determined by the diffusive modes, which feature 
$\Omega_n\sim T \sim q^2$, rather than by the critical ones. This leads to
$[T]=\phi/\nu$, with $\phi$ as given in Eq.\ (\ref{eq:3.9b}). The
proper interpretation of $\phi$ is that of a
crossover exponent associated with the crossover from the quantum to
the thermal fixed point that occurs at any $T>0$. Since $z>\phi/\nu$,
the critical scaling $T\sim b^z$ would be the dominant temperature
dependence if $m$ and $\chi_m$ depended on the critical modes. That they
do not can also been seen from a determination of the magnetic susceptibility
directly from the Gaussian action: Recognizing that the coefficient of
$M^2$ in $\Phi_2$, Eq.\ (\ref{eq:3.1a}), is the inverse spin susceptibility,
and using Eq.\ (\ref{eq:2.12b}) again, we obtain
\begin{equation}
\chi_m(t,T) = {1\over t + T^{1/2\nu}}\quad,
\label{eq:3.10c}
\end{equation}
\end{mathletters}%
in agreement with Eqs.\ (\ref{eq:3.10a},\ \ref{eq:3.10b}).

We are now in a position to determine the exact scale dimension of $u_4$,
and of the other coefficients in the field theory, and to thus verify
the assumption made in the last subsection. As we have seen after 
Eq.\ (\ref{eq:3.6}), the cutoff $q$ scales like $m^{1/2}$, so that
$[q]=[m]_{\rm eff}/2$, where $[m]_{\rm eff}=\beta/\nu$ is the {\it effective}
scale dimension of $m$, i.e. the scale dimension after the effects of
the dangerous irrelevant variable $u_4$ have been taken into account.
From Eqs.\ (\ref{eq:3.8a}), (\ref{eq:3.2a}) we see that $[m]_{\rm eff}\leq 2$,
which justifies the assumption made in Sec.\ \ref{subsec:III.B} that led
to the upper bounds on the scale dimensions of the $u_l$. Repeating the
power counting arguments that led to the inequalities, Eqs.\ (\ref{eq:3.4})
and (\ref{eq:3.5a}), we obtain
\begin{equation}
[u_3] = -(d-2)/2\quad,
\eqnum{3.4'}
\label{eq:3.4'}
\end{equation}
and
\begin{equation}
[u_4] = \cases{-(d-2)           & for $2<d<4$\cr%
               -(d^{2}-12)/(d-2)& for $4<d<6$\cr%
               -d               & for $d>6$\cr}%
         \quad.
\eqnum{3.5a'}
\label{eq:3.5a'}
\end{equation}
The exact scale dimension of the general coefficient $u_l$, if desired, 
can be obtained by the same argument. Again, the upper bound given in
Eq.\ (\ref{eq:3.5c}) is equal to the exact value for $2<d<4$. These
results complete the proof that $d=2$ is the upper critical dimension
for our problem, and that an infinite number of operators become marginal
in $d=2$, and relevant in $d<2$.

Finally, we mention how to reconcile the value for the exponent
$\beta$, Eq.\ (\ref{eq:3.8a}), with scaling. Putting $T=H=0$ for simplicity,
but keeping the dependence of
$m$ on $u_4$ explicitly, and using Eq.\ (\ref{eq:3.3b}), we write
\begin{equation}
m(t,u_4) = b^{-(d+2)/2}\,m(tb^{1/\nu},u_4b^{[u_4]})\quad.
\eqnum{3.9a'}
\label{eq:3.9a'}
\end{equation}
From Eq.\ (\ref{eq:3.6}) we know that $m\sim u_4^{-1/2}$. Using this in
Eq.\ (\ref{eq:3.9a}) changes the scale dimension of $m$ from $[m]=(d+2)/2$ to
an effective value, $[m]_{eff}=[m]-[u_4]/2$. With Eq.\ (\ref{eq:3.5a'}) we
then obtain $[m]_{eff} = \beta/\nu$ with the correct values of $\beta$ and
$\nu$.

\subsection{The specific heat}
\label{subsec:III.D}

The scaling equation for the specific heat is determined by the sum of
the mean-field and the Gaussian fluctuation contribution to the
free energy density, $f$. The mean-field contribution follows immediately from
Eq.\ (\ref{eq:3.7}). The Gaussian fluctuation contribution, $f_G$, 
which gives the leading nonanalyticity at the critical point, can be 
calculated by standard methods.\cite{Ma} Neglecting an uninteresting
constant contribution to $f_G$, we obtain,
\begin{eqnarray}
f_G = {T\over 2V} \sum_{{\bf q},\omega_n} \biggl[2 \ln\left(
       H/m + a_{d-2}\,q^{d-2}\right.
\nonumber\\
       \left. + a_2\,q^2 + a_{\omega}\,\vert\omega_n\vert
       /q^2\right)
\nonumber\\
      + \ln\left(x_d\,H/m - (x_d-1)\,t + a_{d-2}\,q^{d-2}\right.
\nonumber\\
      \left. + a_2\,q^2 + a_{\omega}\,
           \vert\omega_n\vert/q^2\right)\biggr]\quad.
\label{eq:3.11}
\end{eqnarray}
Here $x_d = d/2$ for $2<d<6$, and $x_d=3$ for $d>6$. 
The specific heat coefficient $\gamma_V$ is conventionally defined by,
\begin{equation}
\gamma_V = c_V/T = -\partial^2 f/\partial T^2\quad.
\label{eq:3.12}
\end{equation}
Again we are interested only in scaling properties and not in exact
coefficients. Schematically, Eqs.\ (\ref{eq:3.11}) and (\ref{eq:3.12})
give,
\begin{equation}
\gamma_V = \int_0^{\Lambda} dq\ {q^{d-1}\over T+q^d+q^4+Hq^2/m}\quad,
\label{eq:3.13}
\end{equation}
with $\Lambda$ an ultraviolet cutoff.

Several points should be noted. First, for all $d$ that obey $2<d<4$,
$\gamma_V$ given by Eq.\ (\ref{eq:3.13}), or exactly by Eqs.\ (\ref{eq:3.11})
and (\ref{eq:3.12}), is logarithmically singular for $T,H\rightarrow 0$.
This $d$-independent logarithmic singularity is somewhat unusual.
Wegner has discussed how logarithmic corrections to scaling arise
if a set of scale dimensions fulfills some resonance condition.\cite{Wegner}
In the present case the appearance of a logarithm can be traced to the
fact that the scale dimension of the free energy, $d+z=2z$ for $2<d<4$,
is a multiple of the scale dimension of $T$, which is $z=d$ in this region.
The unusual feature of the logarithm appearing in a {\it range} of
dimensions, rather than only for a special value of $d$, is due to the
dynamical exponent being exactly $d$ in that range.
Second, in Eq.\ (\ref{eq:3.13}) two different temperature scales appear.
The first two terms in the denominator indicate that $T\sim\xi^{-d}$, as
one would expect from dynamical scaling. However, the last term in
Eq.\ (\ref{eq:3.13}) contains the magnetization, which in turn depends
on the crossover temperature scale, $T\sim\xi^{-2}$, see
Sec.\ \ref{subsec:III.C} above. These two features
imply that the scaling equation for $\gamma_V$ should be written,
\begin{eqnarray}
\gamma_V(t,T,H) = \Theta(4-d)\,\ln b \qquad\qquad\qquad\qquad
\nonumber\\
            + F_{\gamma}(t\,b^{1/\nu},T\,b^z,
                  H\,b^{\delta\beta/\nu},T\,b^2)\quad.
\label{eq:3.14}
\end{eqnarray}
Since $z>2$, one can formally ignore the fourth entry in the scaling
function since it is subleading compared to the second entry. The
corrections to the resulting theory can be considered as `corrections
to scaling'. Notice that in contrast to the magnetization and the magnetic
susceptibility, the specific heat does depend on the critical modes, and
hence contains the critical temperature scale. As mentioned in the last
subsection, the latter is dominant when it is present, and $\gamma_V$
provides an example for that.

\section{Discussion}
\label{sec:IV}

\subsection{Theoretical aspects}
\label{subsec:IV.A}

In this paper we have shown that in disordered itinerant quantum
ferromagnets the diffusive nature of the electrons leads to long-range
interactions between spin fluctuations in an order parameter field
theory. As for classical models with long-range 
interactions,\cite{FisherMaNickel} the critical behavior of this field
theory can be determined exactly. For $d>6$, standard mean-field results
are obtained, but for $2<d<6$ one finds nontrivial critical behavior
with dimensionality dependent exponents. For $d\leq 2$ our approach
breaks down because the electrons are localized, and because in our field
theory an infinite number of operators become marginal in $d=2$. The
exact critical exponents for $d>2$ are given by Eqs.\ (\ref{eqs:3.2}), 
(\ref{eqs:3.8}), and (\ref{eq:3.10b}), and the scaling properties of some
of the more interesting physical properties are given by Eqs.\ 
(\ref{eqs:3.9}), (\ref{eq:3.10a}), and (\ref{eq:3.14}).
In this subsection we discuss various aspects of these results that have
not been covered yet.

First of all, there
is an important conceptual question that should be considered. In
our approach we have assumed that it is sensible to construct an order
parameter field theory to describe the critical behavior of the order
parameter. In general this procedure will break down if there are
other soft or slow modes that couple to the order parameter fluctuations.
That is, a more complete low energy theory might be needed. 
In the present case, the diffusons that lead to the 
nonanalyticities in the bare field theory and to, e.g.,
Eqs.\ (\ref{eqs:2.12}) and (\ref{eqs:2.15}) are such soft modes. 
One should ask why we were able to proceed with an order parameter
description anyway, without running into unsurmountable difficulties
due to the additional soft modes. A technical answer is that the
diffusons did create problems, namely divergent coefficients in the
LGW functional, but that for the present problem these difficulties
could be dealt with. Nevertheless, one might wonder what the theory
would look like if the critical modes and soft diffusion modes were treated
on a more equal footing.

Without realizing it, we have previously addressed the above question.
Early work on the metal-insulator transition in disordered interacting
electron systems showed that in two-dimensional systems without impurity
spin-flip scattering, the triplet interaction scaled to large values
under renormalization group iterations.\cite{RunawayFlow} This was
interpreted, incorrectly as it turned out later, as a signature of local
moment formation in all dimensions.\cite{LM} 
Subsequently, the present authors
studied this problem in some detail.\cite{IFS} We were able to explicitly
resum the perturbation theory and show that at a critical value of the
interaction strength, or of the disorder, there is a bulk, thermodynamic
phase transition. The physical meaning of this phase transition was obscure
at the time since no order parameter had been identified, and its description 
was entirely in terms of soft diffusion modes. However, the critical exponents
obtained are identical to those obtained here for the quantum ferromagnetic
phase transition, and in both cases logarithmic corrections to scaling
are found. Because the exponents in the two cases are identical, we
conclude that the transition found earlier by us, whose physical nature
was unclear, is actually the ferromagnetic transition. One also concludes 
that our speculations about the nature of the ordered phase as an 
`incompletely frozen spin phase' with no long-range magnetic order, were
not correct. On the other hand, the techniques used in Ref.\ \onlinecite{IFS}
allowed for a determination of the qualitative phase diagram as a function
of dimensionality, which our present analysis is not capable of. The 
theory given
here not only explains the nature of the transition, but also explains
why the critical behavior at that phase transition could be obtained
exactly in three dimensions: The long-range nature of the interactions
between the order parameter fluctuations makes the critical phenomena
problem exactly soluble.
It is also interesting to note that the list of scaling
scenarios for soft-mode field theories for disordered interacting fermions
given in Sec.\ IV of Ref.\ \onlinecite{R} included the present case, namely
a transition to a ferromagnetic state with an order parameter exponent
$\beta=2\nu$ for $d<4$.

It should also be pointed out that our earlier theory depended crucially 
on there being electronic spin conservation. This feature would be lost
of there were some type of impurity  spin flip scattering process. In
that case, the soft modes that lead to the long-range order parameter
interactions acquire a mass or energy gap, and at sufficiently large
scales the interactions are effectively of short range. The asymptotic
critical phenomena in this case are described by a short-range, local
order parameter field theory with a random mass, or temperature, term.
In this case the random mass term {\it is} relevant with respect to
the Gaussian fixed point analogous to the one discussed here, which
underscores the important role that is played by the order parameter
being conserved in our model. The quantum phase transition in a model 
where it is not, and where the random mass term is therefore relevant with
respect to the analogue of our Gaussian fixed point, is discussed 
elsewhere.\cite{afm}  We also mention that the effect of fermionic
soft modes on the ferromagnetic phase transition in {\it clean} systems
has been discussed recently in Ref.\ \onlinecite{clean}.

Even though we quote exponent values formally for all $d>2$, it should
be mentioned that the region of validity of our analysis shrinks to zero
as $d$ approaches $2$.
As mentioned in Sec.\ \ref{subsec:II.C}, the reference system has
all of the characteristics of the system described by the full action $S$,
except that it must not undergo a phase transition lest our separation of
modes that is implicit in our singling out $S_{\rm int}^t$ for the
decoupling procedure break down.
This requirement puts restrictions on the parameter
values for which our procedure works. For instance, we cannot go to
arbitrarily large disorder (at fixed $d$) without triggering a metal-insulator
transition within the reference system. For $d\rightarrow 2$ the 
metal-insulator transitions occurs at smaller and smaller values of the
disorder, and in $d=2$ one obtains a very complicated, and unsolved,
situation where various fluctuations compete with each other.

We finally discuss why some of our results are in disagreement with
Sachdev's\cite{Sachdev} recent general scaling analysis of quantum phase 
transitions with conserved order parameters. For instance, it follows from
our Eqs.\ (\ref{eqs:3.10},\ \ref{eq:3.14}) that the Wilson ratio, defined as
$W = (m/H)/(C_V/T)$, diverges at criticality rather than being a universal
number as predicted in Ref.\ \onlinecite{Sachdev}. Also, for $2<d<4$ the
function $F_{\gamma}$ in Eq.\ (\ref{eq:3.14}), for $t=0$ and neglecting
corrections to scaling, is a function of $T/H$, in agreement with
Ref.\ \onlinecite{Sachdev}, but for $d>4$ this is not the case. 
The general reason for this breakdown of general scaling is that we work
above an upper critical dimensionality, and hence dangerous irrelevant
variables have to be considered very carefully, and on a case by case
basis. This caveat is particularly relevant for quantum phase transitions
since they tend to have a low upper critical dimension.
It is well known that a given irrelevant variable can be
dangerous with respect to some observables but not with respect to others.
Specifically, in our case the dangerously irrelevant variable
$u_4$ affects the leading scaling behavior of the magnetization,
but not that of the specific heat coefficient, which leads to the divergence
of the Wilson ratio. A simple example of this phenomenon 
is provided by classical $\phi^4$-theory in $d>4$, where the dangerous 
irrelevant variable $u$ (the coefficient of the $\phi^4$-term) affects
the scale dimension of the magnetization, but not that of the specific 
heat.\cite{Ma} In classical $\phi^4$-theory this point is
obscured by the fact that the saddle-point contribution to the specific
heat contains a discontinuity. This is often expressed as $\alpha=0$, with
$\alpha$ the specific heat exponent. However, the approach to the
discontinuity is described by the $\alpha$ suggested by hyperscaling,
namely $\alpha = 2-d/2$, see ch. VII.4 of Ref.\ \onlinecite{Ma}. At the quantum
FP the situation is clearer, since the saddle-point
contribution to $C_V$ is subleading. It is also important to remember that
different arguments of a scaling function can be affected in different ways
by one and the same dangerous irrelevant variable. Here, the effective scale
dimension of $H$ in the specific heat is changed by $u_4$ (from $(3d-2)/2$
to $d$ in $2<d<4$), but that of $T$ is not, since $u_4$ imports only the
subleading crossover temperature scale into $\gamma_V$ via the appearance
of $m$ in Eq.\ (\ref{eq:3.13}).

\subsection{Experimental aspects}
\label{subsec:IV.B}

\begin{figure}
\epsfxsize=8.25cm
\epsfysize=6.6cm
\epsffile{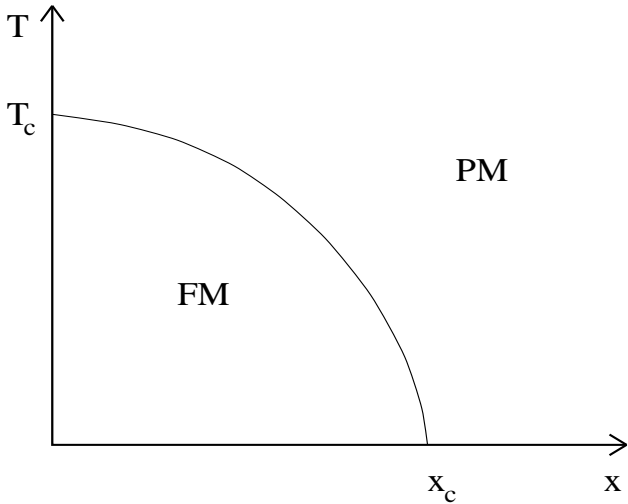}
\vskip 0.5cm
\caption{Schematic phase diagram for an alloy of the form 
${\rm P}_x\,{\rm F}_{1-x}$. $T_c$ is the Curie temperature for the pure
ferromagnet F, and $x_c$ is the critical concentration.}
\label{fig:2}
\end{figure}

In order to apply our theoretical results to experiments, one needs 
materials that show a transition from a paramagnetic state to a ferromagnetic
one at zero temperature as a function of some parameter $x$. Obvious
candidates are magnetic alloys of the form ${\rm P}_x\,{\rm F}_{1-x}$ with
P a paramagnetic metal, and F a ferromagnetic one. Such materials show the
desired transition as a function of the composition parameter $x$; examples
include ${\rm F}={\rm Ni}$ and ${\rm P}={\rm Al,}\ {\rm Ga}$.\cite{Mott} 
The schematic phase diagram is shown in Fig.\ \ref{fig:2}.

One striking difference between our results for the quantum ferromagnetic
phase transition at $(x=x_c,T=0)$ and the classical or thermal transition
for Heisenberg ferromagnets are the numerical values of the exponents.
For $3$-$d$ systems, our Eqs.\ (\ref{eqs:3.8}) for instance predict
$\beta=2$, while the corresponding value for the thermal
transition is,\cite{ZJ} $\beta_{\rm class}\approx 0.37$.
The large difference between the classical
and the quantum value should be easily observable if it was possible to
measure the magnetization at a sufficiently low temperature as a function
of $x$ in order to observe the crossover between quantum and classical
critical behavior in the vicinity of $x_c$. One possible way to do such 
an experiment would involve the preparation of many samples with different 
values of $x$ over a small region of $x$. It might also be possible to probe
the magnetic phase transition by using the stress tuning technique that was 
used to study the metal-insulator transition in Si:P.\cite{stress}
Alternatively, one could prepare a sample with a
value of $x$ that is as close as possible to $x_c$, and measure the 
magnetic field dependence of the magnetization, extrapolated to $T=0$, to
obtain the exponent $\delta$. Again, there is a large difference between
our prediction of $\delta=1.5$ in $d=3$, and the classical value
$\delta_{\rm class}\approx 4.86$.

\begin{figure}
\epsfxsize=8.25cm
\epsfysize=6.6cm
\epsffile{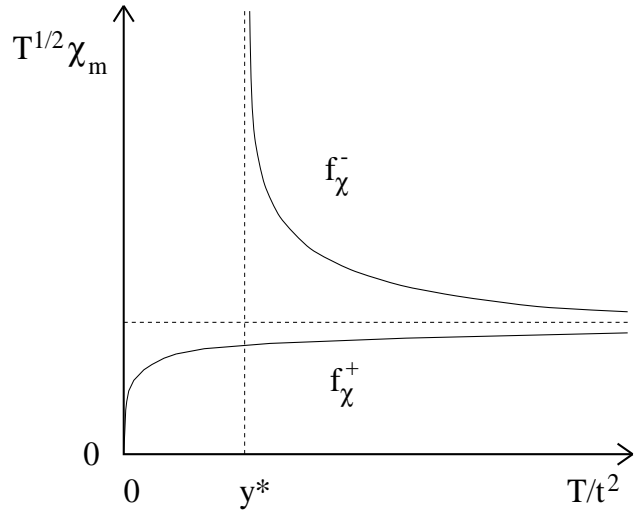}
\vskip 0.5cm
\caption{Schematic prediction for a scaling plot of the magnetic 
suscteptibility.}
\label{fig:3}
\end{figure}

Another possibility, that does not involve an extrapolation to $T=0$,
is to measure the zero-field magnetic susceptibility as a function of both
$t=\vert x-x_c\vert$ and $T$. Equation (\ref{eq:3.10a}) predicts
\begin{equation}
\chi_m(t,T) = T^{-1/2}\,f_{\chi}(T/t^2)\quad.
\label{eq:4.1}
\end{equation}
Here $f_{\chi}$ is a scaling function that has two branches, $f_{\chi}^+$
for $x>x_c$, and $f_{\chi}^-$ for $x<x_c$. Both branches approach a constant 
for large values of their argument, 
$f_{\chi}^{\pm}(y\rightarrow\infty)={\rm const.}$ For small arguments, we
have $f_{\chi}^+(y\rightarrow 0)\sim \sqrt{y}$, while $f_{\chi}^-$ diverges
at a nonzero value $y^*$ of its argument that signalizes the classical
transition, $f_{\chi}^-(y\rightarrow y^*)\sim (y-y^*)^{-\gamma_{\rm class}}$,
with $\gamma_{\rm class}\approx 1.39$ the susceptibility exponent for
the classical transition. Our prediction is then that a plot of
$\chi_m\ T^{1/2}$ versus $T/t^2$ will yield a universal function the
shape of which is schematically shown in Fig.\ \ref{fig:3}.
Notice that the exponents are known {\it exactly}, so the only adjustable
parameter for plotting experimental data will be the position of the
critical point. This is on sharp contrast to some other quantum phase
transitions, especially metal-insulator transitions, where the exponent
values are not even approximately known, which makes scaling plots almost
meaningless.\cite{glass}

Finally, one can consider the low-temperature behavior of the specific
heat. According to Eq.\ (\ref{eq:3.14}), as the temperature is lowered
for $x\agt x_c$ the leading temperature dependence of
the specific heat will be
\begin{mathletters}
\label{eqs:4.2}
\begin{equation}
c_V(T) \sim T\ln T\quad.
\label{eq:4.2a}
\end{equation}
At criticality this behavior will continue to $T=0$, while for $x>x_c$
it will cross over to
\begin{equation}
c_V(T) \sim (\ln t)\ T\quad.
\label{eq:4.2b}
\end{equation}
\end{mathletters}%
For $x\alt x_c$ one will encounter the classical Heisenberg transition
where the specific heat shows a finite cusp (i.e., the exponent $\alpha <0$).

\acknowledgments

We thank the TSRC in Telluride, CO for hospitality during
the 1995 Workshop on Quantum Phase Transitions, where part of this work
was performed, and the participants of the workshop for stimulating
discussions. This work was supported by the NSF under grant numbers 
DMR-92-17496 and DMR-95-10185.

\appendix
\section{The wavenumber dependent spin susceptibility}
\label{app:A}
Here we calculate the wavenumber dependent spin susceptibility
in a disordered interacting Fermi system. In terms of the $Q$-matrix
field theory reviewed in Ref.\ \onlinecite{R}, the spin susceptibility
reads,
\begin{eqnarray}
\chi_s({\bf q},\Omega_n) \sim \sum_{r=0,3} (-)^r\ T\sum_{m_1,m_2}
 \left\langle ^{3}_{r}Q_{m_1+n,m_1}^{\alpha \alpha}({\bf q})\right.
\nonumber\\
 \left.\times ^{3}_{r}Q_{m_2-n,m_2}^{\alpha \alpha}(-{\bf q})\right\rangle\quad.
\label{eq:A.1}
\end{eqnarray}
The classical matrix field $Q$ comprises two fermionic degrees of freedom,
and a general matrix element $^{i}_{r}Q_{nm}^{\alpha\beta}$ has six
indices: $i$ is the spin index (with $i=1,2,3$ the spin triplet channel),
$r=0,3$ denotes the particle-hole channel ($r=1,2$ would be the 
particle-particle channel), $n,m$ are Matsubata frequency indices, and
$\alpha,\ \beta$ are replica indices. The matrix elements with 
$nm<0$ describe the soft modes, i.e., the particle-hole excitations, while
those with $nm>0$ are massive. For explicit calculations it is
convenient to consider the nonlinear $\sigma$-model version of the
$Q$-field theory,\cite{Finkelstein,R} where the massive modes are integrated
out. In the resulting effective model, $Q^2=1$, and a loop expansion can be
set up by expanding all $Q$ in terms of $q_{nm}=\Theta(-nm)\,Q_{nm}$. The
contribution of $O(q^2)$ to Eq.\ (\ref{eq:A.1}) vanishes for $\Omega_n=0$
and hence does not contribute to 
$\chi_0({\bf q}) = \chi_s({\bf q},\Omega_n=0)$. The one-loop contribution,
which is the term of $O(q^4)$, gives
\begin{mathletters}
\label{eqs:A.2}
\begin{eqnarray}
\chi_0({\bf q}) \sim {1\over V}\sum_{\bf k} 
   T\sum_{n} \omega_n\,{\cal D}_n^t({\bf k})\,{\cal D}_n({\bf k})\,
\nonumber\\
     \times\left[{\cal D}_n({\bf k}-{\bf q}) - 
                 {\cal D}_n^t({\bf k}-{\bf q})\right]\quad.
\label{eq:A.2a}
\end{eqnarray}
Here ${\cal D}_n$ and ${\cal D}_n^t$ are the diffusive propagators of the
theory.\cite{R} Their structure is
\begin{equation}
{\cal D}_n({\bf k}) = 1/(k^2 + D^{-1}\omega_n)\quad,
\label{eq:A.2b}
\end{equation}
\end{mathletters}%
with $D$ a diffusion coefficient. ${\cal D}_n^t$ has the same structure,
with $D$ replaced by $D^t\neq D$. Since we are not interested in prefactors,
we do not have to specify either $D$ and $D^t$, or the prefactor in
Eq.\ (\ref{eq:A.2a}). For the reasons discussed at the end of 
Sec.\ \ref{subsec:II.C}, the one-loop term suffices to calculate the
leading infrared wavenumber dependence of $\chi_0$. Schematically,
Eqs.\ (\ref{eqs:A.2}) yield,
\begin{equation}
\chi_0({\bf q}) \sim \int_q^{\Lambda} dk\ k^{d-1}\int_0^{\infty}d\omega\ 
        {\omega\over (k^2 + \omega)^3}\quad,
\label{eq:A.3}
\end{equation}
with $\Lambda$ an ultraviolet cutoff,
from which one readily obtains Eq.\ (\ref{eq:2.12a}).

\section{Logarithmic corrections to scaling in $d=4$ and $d=6$}
\label{app:B}

There are three distinct mechanisms that produce logarithmic corrections
to scaling: (1) Marginal operators, (2) Wegner resonance conditions
between a set of scale dimensions, and (3) logarithmic corrections to
the scale dimension of a dangerous irrelevant operator. The first two
mechanism are well known.\cite{Wegner} The third is operative only above
an upper critical dimension, and is therefore of particular interest for
quantum phase transitions.

In the present case, logarithmic corrections to scaling arise due to all
three of these mechanisms. The second one produces corrections to the
scaling of the specific heat in all dimensions $2<d<4$, as was discussed
in Sec.\ \ref{subsec:III.D}. The first one is operative in $d=4$, where $v_4$
is a marginal operator, see Eq.\ (\ref{eq:3.5b}). If desired, the resulting
corrections to scaling can be worked out using standard 
techniques.\cite{Wegner} Finally, the third mechanism produces corrections
to scaling in $d=6$. According to Eq.\ (\ref{eq:2.9}), the coefficient
$u_4 \sim \ln q$ in $d=6$. Via Eq.\ (\ref{eq:3.6}) or (\ref{eq:3.7}) this
leads, for instance, to a leading behavior of the spontaneous magnetization,
\begin{equation}
m(t,H=0) \sim {t^{1/2}\over \sqrt{\ln(1/t)}}\quad,
\label{eq:B.1}
\end{equation}
and at the critical point we have
\begin{equation}
m(t=0,H) \sim {H^{1/3} \over \left[\ln(1/H)\right]^{1/3}}\quad.
\label{eq:B.2}
\end{equation}
Other consequences, e.g. for the specific heat in a magnetic field, can
be easily worked out.


\begin{references}
\bibitem{WeakLoc} The term `weak localization' is ill-defined, and often
 misunderstood. Here we use it to refer to the
 nonanalytic behavior of electronic correlation functions in the limit
 of zero momentum and/or frequency that is induced by quenched disorder,
 or by a combination of interactions and quenched disorder, and that
 occurs even if the system is far from any kind of metal-insulator
 transition. The physical reason for these nonanalyticities is the diffusive
 motion of the electrons in the presence of quenched disorder; for a 
 detailed discussion see, e.g., Refs.\ \onlinecite{LeeRama} and \onlinecite{R}.
\bibitem{connfootnote} This is a consequence if this correlation function
 being {\it disconnected} with respect to the quantum mechanical average.
 Notice that it is a function of three positions, though, since it is 
 {\it connected} with respect to the disorder average.
\end{references}
\end{document}